# Earth-like and Tardigrade survey of exoplanets

MadhuKashyap Jagadeesh[1*], Milena Roszkowska[2,3] and Łukasz Kaczmarek[2]

[1]Department of Physics, Jyoti Nivas College, Bengaluru-560095, Karnataka, India, e mail: kas7890.astro@gmail.com

[2]Department of Animal Taxonomy and Ecology, Faculty of Biology, Adam Mickiewicz University, Poznań, Umultowska 89, 61-614 Poznań, Poland, e-mail: kaczmar@amu.edu.pl

[3]Department of Bioenergetics, Faculty of Biology, Adam Mickiewicz University, Poznań, Umultowska 89, 61-614 Poznań, Poland, e-mail: mil.roszkowska@gmail.com

*Corresponding author

**Abstract**

Finding life on other worlds is a fascinating area of astrobiology and planetary sciences. Presently, over 3500 exoplanets, representing a very wide range of physical and chemical environments, are known. Scientists are not only looking for traces of life outside Earth, but are also trying to find out which of Earth's known organisms would be able to survive on other planets. Tardigrades (water bears) are microscopic invertebrates that inhabit almost all terrestrial, freshwater and marine habitats, from the highest mountains to the deepest oceans. Thanks to their ability to live in a state of cryptobiosis, which is known to be an adaptation to unpredictably fluctuating environmental conditions, these organisms are able to survive when conditions are not suitable for active life; consequently, tardigrades are known as the toughest animals on Earth. In their cryptobiotic state, they can survive extreme conditions, such as temperatures below -250°C and up to 150°C, high doses of ultraviolet and ionising radiation, up to 30 years without liquid water, low and high atmospheric pressure, and exposure to many toxic chemicals. Active tardigrades are also resistant to a wide range of unfavourable environmental conditions, which makes them an excellent model organism for astrobiological studies. In our study, we have established a metric tool for distinguishing the potential survivability of active and cryptobiotic tardigrades on rocky-water and water-gas planets in our solar system and exoplanets, taking into consideration the geometrical means of surface temperature and surface pressure of the considered planets. The Active Tardigrade Index (ATI) and Cryobiotic Tardigrade Index (CTI) are two metric indices with minimum value 0 (= tardigrades cannot survive) and maximum 1 (= tardigrades will survive in their respective state). Values between 0 and 1 indicate a percentage chance of the active or cryptobiotic tardigrades surviving on a given exoplanet. Among known planets (except for Earth), the highest values of the ATI and CTI indices were calculated for the exoplanets Mars Kepler-100d, Kepler-48d, Kepler-289b, TRAPPIST-1 f and Kepler-106e.

## 1. Introduction

Exploring the unknown worlds outside our solar system (i.e., exoplanets or moons) is part of a new era in current research. The systematic scientific approach in this direction was begun in 1992, by Aleksander Wolszczan and Dale Frail, by observing the planets orbiting around pulsating stars. In 1995, the first confirmed exoplanet (planet outside of our solar system that orbits a star)—51 Pegasi b, orbiting around the 51 Pegasi G-type star—was discovered (Mayor &Queloz 1995). Since then, many exoplanets have been detected, using methods such as radial velocity, transit, gravitational microlensing and direct imaging (Sengupta 2016). New techniques have allowed the discovery of extrasolar planets in huge numbers per year. The former Kepler mission (2009–2013), and present (second) Kepler mission, K2 (active since 2014), were designed to survey the region encompassed by the Milky Way galaxy to discover Earth-sized and smaller planets, in or near the habitable zone of Sun-like stars, and to determine the solar systems in our galaxy that might have such planets (https://www.nasa.gov). All such planets could, in future years, be considered for exploration for signs of life. Since 2009, when the Kepler mission started, the number of confirmed exoplanets increased greatly. Presently, more than 3500 exoplanets have been confirmed (from about 300 known before the Kepler mission), according to the Planetary Habitability Laboratory PHL-HEC (http://phl.upr.edu/projects/habitable-exoplanets-catalog/data/database), which is maintained by the University of Puerto Rico, Arecibo.

Indexing is a primary criterion used to give structure to these raw data from space missions such as CoRoT and Kepler. There have been approximately more than 1650 rocky exoplanets confirmed within a radius of 0.5 to 1.9 Earth units (EU), and with a mass of 0.1 to 10 EU (Kashyap et al. 2017). A few years ago, Schulze-Makuch et al. (2011) introduced a physical and chemical parameters-based index of exoplanets, in the forms of an Earth Similarity Index (ESI) and a Planetary Habitability Index (PHI). The ESI is defined as a geometrical mean of four physical parameters (such as radius, density, escape velocity and surface temperature), the metric ranges of which are between 0 (dissimilar to Earth) to 1 (similar to Earth). The PHI approach requires the presence of a stable substrate, with appropriate chemistry, that could hold a liquid solvent, which could support life; however, to determine if such a stable substrate having these features is present on all exoplanets is, at present, a challenging task.

Similarly, in this paper, the Active Tardigrade Indexand cryptobioticTardigrade Indexare introduced and re-defined as the geometrical mean of seven physical parameters (radius, density, escape velocity, surface temperature, surface pressure, planetary revolution, surface gravity). Since planetary revolution is not directly available as raw data, the values have been calculated for 60 (as of August 2017) confirmed exoplanets with a water medium.

The phylum Tardigrada (water bears) consists of over 1200 species (Guidetti&Bertolani 2005; Degma&Guidetti 2007; Degma et al. 2009–2017) that inhabit aquatic (freshwater and marine) and terrestrial environments throughout the world, from the deepest seas to the highest mountains (Ramazzotti&Maucci 1983; McInnes 1994; Nelson et al. 2010, 2015). Water bears are one of the toughest metazoans on Earth, and are often used in research on survivability in extreme conditions (Wright 2001; Guidetti at al. 2012). It has been shown that many tardigrade species have significant resistance to many physical and chemical environmental stressors, including desiccation, very high and extremely low temperatures, high levels of ionising radiation, high pressures, and chemicals, such as ethanol, carbon dioxide, hydrogen sulphide, 1-hexanol, or methyl bromide gas (e.g., Baumann 1922; Ramløv&Westh 1992, 2001; Seki & Toyoshima 1998; Jönsson et al. 2001, 2007; Horikawa et al. 2006; Ono et al. 2008; Wełnicz et al. 2011; Guidetti et al. 2012). Tardigrades are also able to survive in open space, with combined exposure to the space vacuum and solar and cosmic radiation (e.g. Jönssonet al. 2008). They owe this remarkable resistance to extreme conditions to their ability to enter a reversible physiological state, i.e., cryptobiosis. In cryptobiosis, metabolic processes significantly decrease (Pigoń&Węglarska 1955; Clegg 1973). Entering cryptobiosis, and then returning to active life, requires preparation, which includes: a) forming a tun (Baumann 1922); and b) synthesising many different molecules and bioprotectants, such as non-reducing sugars (e.g., trehalose), late embryogenesis abundant proteins, heat shock proteins, cytoplasmic abundant heat-soluble proteins, secretory abundant heat-soluble proteins, and aquaporin proteins (e.g., Hengherr et al. 2008; Förster et al. 2009; Wełnicz et al. 2011; Guidetti et al. 2011, 2012; Yamaguchi et al. 2012; Grohme et al. 2013). Tardigrades also have very efficient DNA repair mechanisms (e.g., Rizzo et al. 2010; Wełnicz et al. 2011). All of these remarkable abilities make them a perfect multicellular organismal model for astrobiological studies (Jönsson 2007; Horikawa 2008).

In this paper we introduce two indexes i.e. Active Tardigrade Index (ATI) and Cryptobiotic Tardigrade Index (CTI) for 60 rocky-water or water-gas exoplanets and based on these indexes we calculate a probability of surviving of active and cryptobiotic tardigrades on these planets.

## 2. Active Tardigrade Index (ATI)

According to the recent derivation of Kashyap et al. (2017), the global ATI can be mathematically represented as:

$$ATI_x = \left\{1 - \frac{|x-x_0|}{|x+x_0|}\right\}^{w_x} \qquad (1)$$

where x is the physical parameter of the exoplanet, such as radius (R), bulk density (ρ), escape velocity (Ve), planetary revolution (Re), surface gravity, surface pressure, and surface temperature (Ts), whereas x0 is with reference to Earth, and $w_x$ is the weight exponent (see Table 1). These parameters are expressed as EU, while the surface temperature is represented in Kelvin (K).

The weight exponents for the upper and lower limits of the parameters were calculated following Schulze-Makuch et al. (2011): radius 0.5 to 1.9 EU, mass 0.1 to 10 EU, density 0.7 to 1.5 EU, escape velocity 0.4 to 1.4 EU, surface temperature and pressure (T = 1 to 38°C and P = up to 7500 Mpa for active, and T = -253 to 153°C and P = up to 7500 Mpa for cryptobiotic tardigrades). Similarly, we defined the limits of gravity as 0.16 to 17 EU, and planetary revolutions as 0.61 to 1.88 EU. Gravity and planetary revolutions are newly-introduced weight exponents, which can be calculated from the following conditions: the human centrifuge experiment (Brent et al. 2012) clearly showed that untrained humans can tolerate 17 EU, with eyeballs In 2016, Hashimoto et al.discovered that tardigrade protein could be used on human DNA to obtain protection from radiation, which is a key idea to target tardigrade survival on exoplanets. The planetary revolution is scaled on the basis of habitable zone of Sun-like stars, in terms of Earth years.

**Table 1.** The parameters for Active Tardigada Index (ATI) and Cryptobiotic Tardigrade Index (CTI) scale.

| Planetary property | Reference values for ATI and CTI | Weight exponents for ATI | Weight exponents for CTI |
|---|---|---|---|
| Mean radius (R) | 1 EU | 0.57 | 0.57 |
| Bulk density (ρ) | 1 EU | 1.07 | 1.07 |
| Escape velocity ($V_e$) | 1 EU | 0.70 | 0.70 |
| Surface gravity (g) | 1EU | 0.13 | 0.13 |
| Revolution (Re) | 1EU | 0.70 | 0.70 |
| Surface Pressure (P) | 1EU | 0.32 | 0.32 |
| Surface temperature ($T_s$) | 288 K | 7.33 | 0.94 |

The global ATI is divided into interior (ATII) and surface (ATIS), which are expressed as:

$$ATI_I = (ATI_R \times ATI_\rho)^{1/2} \qquad (2)$$

$$ATI_S = (ATI_{Ve} \times ATI_{Ts} \times ATI_P \times ATI_{Rev} \times ATI_g)^{1/5} \qquad (3)$$

Therefore, the global ATI is given as:

$$ATI = (ATI_I \times ATI_S)^{1/2} \qquad (4)$$

Graphically

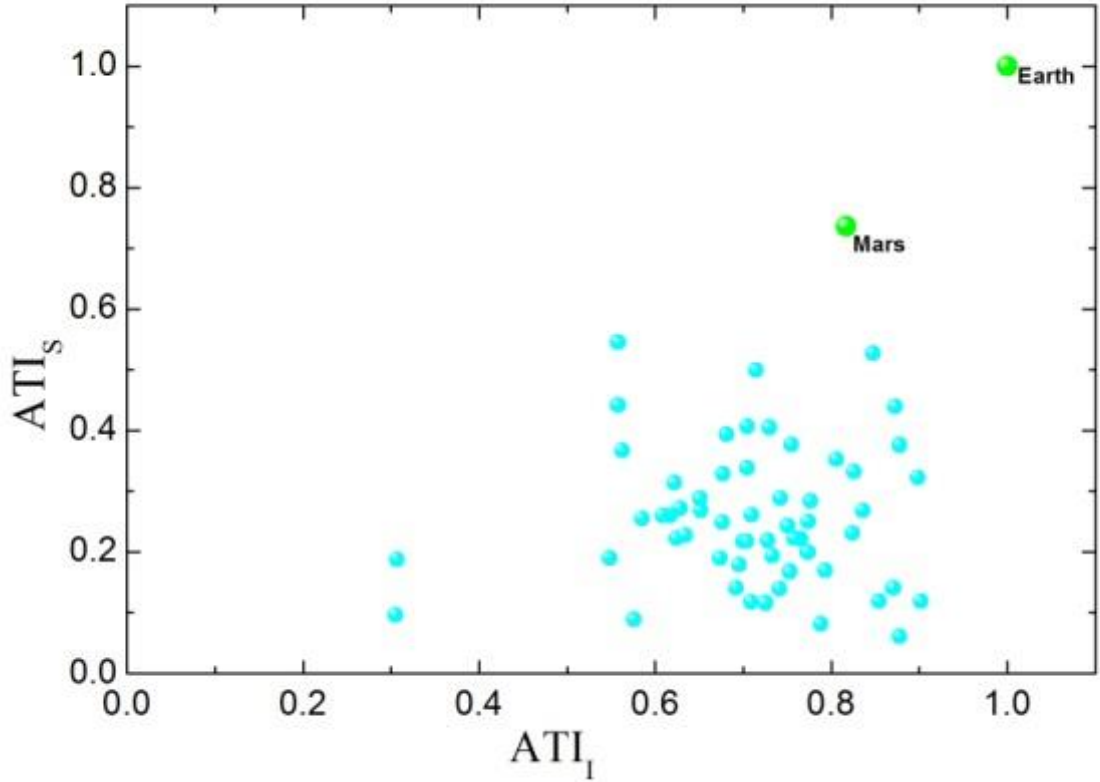

**Figure 1**. Scatter plot of interior(i.e., geometrical mean of radius and density of the planet) and surface (i.e., geometrical mean of surface gravity, pressure, temperature and escape velocity of the planet) ActiveTardigada Index (ATI) (the green dots represent the solar-system objects and blue dots represent the rocky-water and water-gas exoplanets). This graph defines the threshold value of ATI=0.57 (Mars as reference from its ESI value) for tardigrades to survive in Active state.

## 3. Cryptobiotic Tardigrade Index (CTI)

The global CTI can be mathematically represented as:

$$CTI_x = \left\{1 - \frac{|x-x_0|}{|x+x_0|}\right\}^{w_x} \qquad (5)$$

where x is the physical parameter of the exoplanet, such as radius (R), bulk density (ρ), escape velocity (Ve), planetary revolution (Re), surface gravity, surface pressure, and surface temperature (Ts), whereas x0 is with reference to Earth, and $w_x$ is the weight exponent (see Table 1). These parameters are expressed in EU, while the surface temperature is represented in K.

The global CTI is divided into interior (CTII) and surface (CTIS), which are expressed as:

$$CTI_I = (CTI_R \times CTI_\rho)^{1/2} \qquad (6)$$

$$CTI_S = (CTI_{Ve} \times CTI_{Ts} \times CTI_P \times CTI_{Rev} \times CTI_g)^{1/5} \qquad (7)$$

Therefore, the global CTI is given as:

$$CTI = (CTI_I \times CTI_S)^{1/2} \qquad (8)$$

Graphically

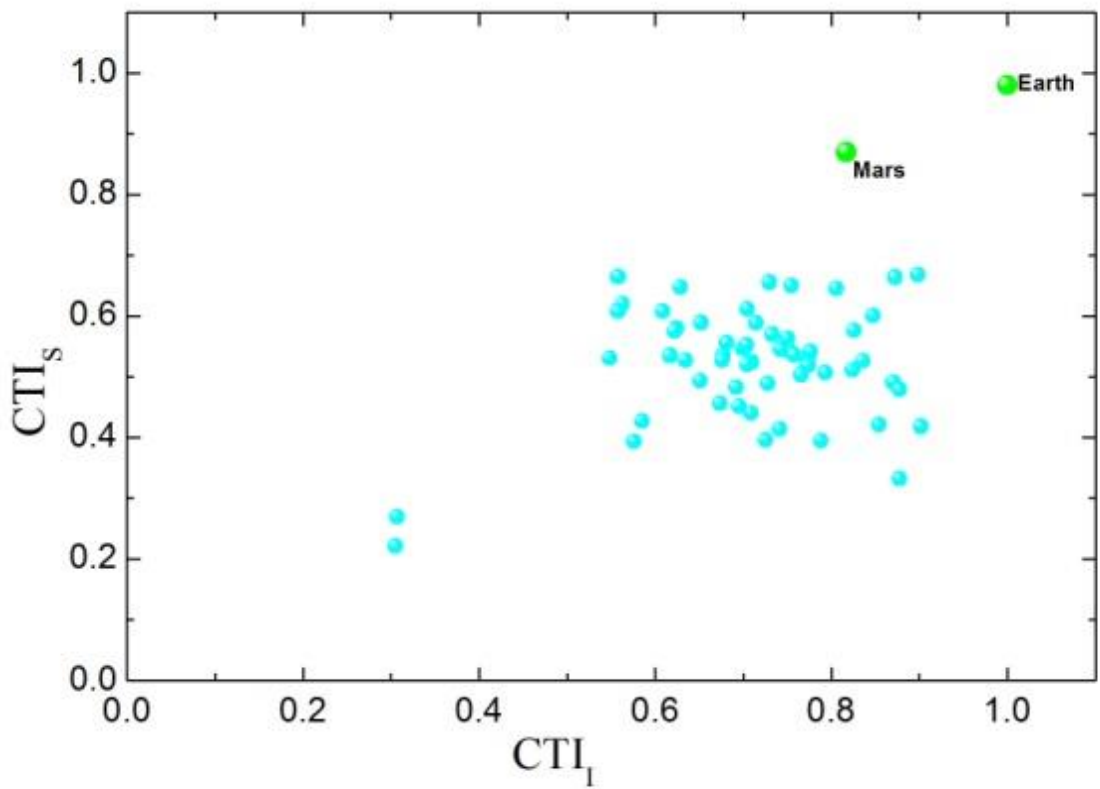

**Figure 2.** Scatter plot of interior (i.e., geometrical mean of radius and density of the planet) and surface (i.e., geometrical mean of surface gravity, pressure, temperature and escape velocity of the planet) Cryptobiotic Tardigrade Index(CTI)(the green dots represent the solar-system objects and blue dots represent the rocky-water and water-gas exoplanets). This graph defines the threshold value of $CTI_s=0.84$(Mars as reference from its ESI value) for tardigrades to survive in cryptobiotic state.

**Table 2.** List of planets with highest Earth Similarity Index (ESI), Active TardigradeIndex (ATI) and Cryptobiotic Tardigrade Index (CTI)with comparison to Earth (R = radius, ρ = density, g = surface gravity, $V_e$= escape velocity, T = surface temperatures, P = surface pressure, Re = revolution, EU = Earth unit, K = kelvin, EY = Earth year).

| Planet Names | R (EU) | ρ (EU) | g (EU) | $V_e$ (EU) | T (K) | P (EU) | Re (EY) | ESI | ATI | CTI |
|---|---|---|---|---|---|---|---|---|---|---|
| Earth | 1 | 1 | 1 | 1 | 288 | 1 | 1 | **1** | **1** | **1** |
| Mars | 0.53 | 0.71 | 0.37 | 0.45 | 218 | 0.99 | 1.88 | **0.73** | **0.57** | **0.84** |
| Kepler-100d | 1.51 | 0.86 | 1.31 | 1.41 | 708.3 | 2.6 | 0.08 | **0.3** | **0.53** | **0.77** |
| Kepler-48d | 2.04 | 0.94 | 1.91 | 1.97 | 493.3 | 7.5 | 0.11 | **0.38** | **0.61** | **0.76** |
| Kepler-289b | 2.15 | 0.74 | 1.58 | 1.84 | 617.5 | 5.4 | 0.09 | **0.3** | **0.53** | **0.72** |
| TRAPPIST-1f | 1.04 | 0.59 | 0.61 | 0.8 | 229.7 | 0.4 | 0.02 | **0.68** | **0.66** | **0.71** |
| Kepler-106e | 2.55 | 0.67 | 1.71 | 2.09 | 576.7 | 7.5 | 0.11 | **0.29** | **0.53** | **0.70** |

The full data are available on [http://dx.doi.org/10.17632/d3fp4wsmx3.1Kashyap and Kaczmarek (2017)].

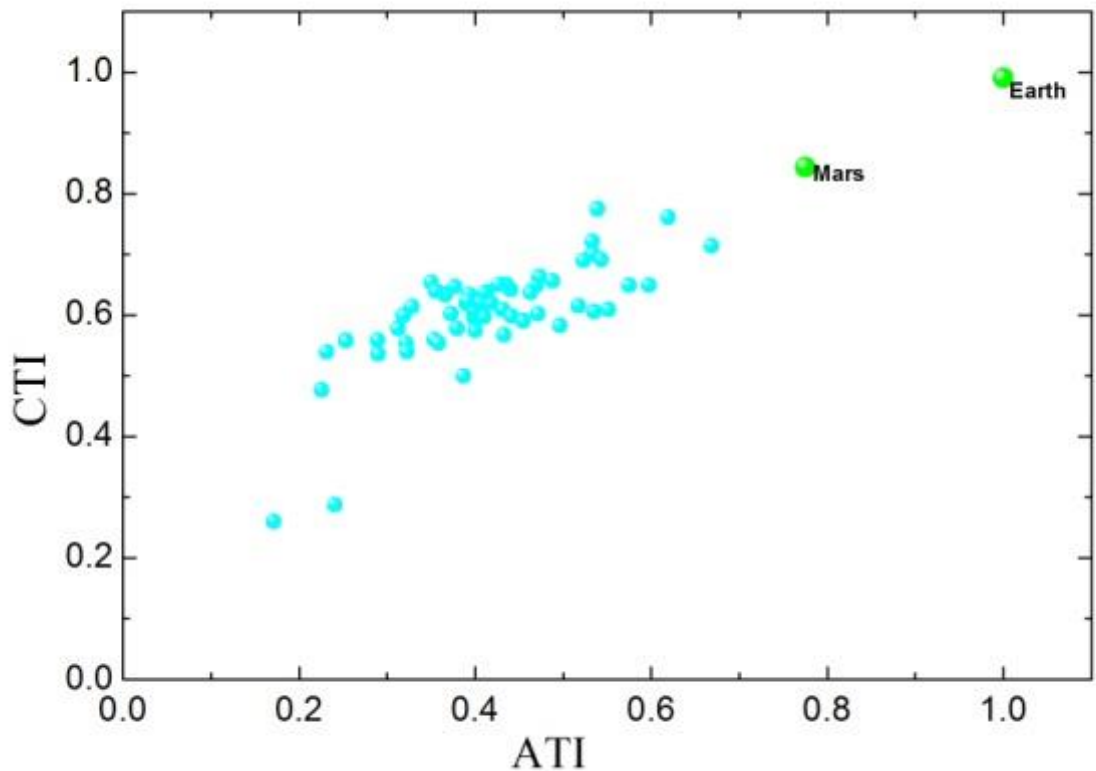

**Figure 3.** Scatter plot of Active TardigradeIndex (ATI) and Cryptobiotic Tardigrade Index (CTI)(the green dots represent the solar-system objects and blue dots represent the rocky-water and water-gas exoplanets).This graph is a representation of CTI and ATI values.

## 4. Conclusions

Searching for traces of life elsewhere in the universe is a challenging and fascinating area of modern astrobiology. In 2011, we started identifying Earth-like planets using metric tools (Schulze-Makuch et al. 2011). Schulze-Makuch et al. (2011) proposed certain physical and chemical parameters to determine how much an exoplanet is similar to Earth, and how high the probability is of finding life on such a planet. In our work, we expanded this idea, choosing rocky-water and water-gas exoplanets and testing the probability of survival on these planets of tardigrades in their active and cryptobiotic stages. For this purpose, we used parameters, such as mean radius, bulk density, escape velocity, surface gravity, planetary revolution, surface pressure and surface temperature.

As has been demonstrated, these tiny animals in their cryptobiotic state are capable of surviving in open space (e.g., Guidetti et al. 2012). From this point of view, tardigrades are proposed as the perfect animals for further astrobiological studies (Erdmann & Kaczmarek 2017). We know the environmental conditions that are suitable for tardigrades living on Earth (ATI and CTI = 1) and, based on the physical and chemical environments that we know are present on Mars, they can probably also survive on this planet (ATI and CTI = 0.77 and 0.84, respectively). This means that similar ATI and CTI values, in the context of other planets, probably indicates that tardigrades could survive on them. At present, it is hard to say if tardigrades could also survive if the physical and chemical parameters were much less favourable than those found on Mars. In addition to tardigrades, some Archaea, bacteria, rotifers, nematodes, lichens and algae are also capable of surviving under extraterrestrial conditions (e.g., Seckbach 2000; Sancho et al. 2007; Borgonie et al. 2011; Rampelotto 2013; Leuko et al. 2014; Oren 2014). This suggests that simple ecosystems, based on these organisms, could theoretically function on some exoplanets, only needing favourable physical and chemical conditions, and some liquid water.

In the indices, we proposed two new parameters (surface gravity and planetary revolution) that were added to a previously existing ESI metric tool for rocky-water and water-gas composition exoplanets only. Based on this, the lowest values of CTI were found for exoplanets such as K2-33 b (CTI = 0.25) with a water-gas composition, and K2-95 b (CTI = 0.28) with a rocky-water composition. The proposed metric tools will potentially highlight which planets, based on physical properties, we can expect to find life, or even simple Earth-like ecosystems. Such knowledge will be very useful for planning further, more detailed observations of preliminarily-selected (based on our indices) space objects, where we might expect Earth-like life.

Exoplanets with values just below the threshold values have a much cooler host star than our sun. In the stellar classification (i.e., O, B, A, F, G, K, M), the host stars fall mostly into K-type or M-type, which are cooler than our sun (G-type), e.g., the TRAPPIST-1f water-system has a M-type host star (https://www.universetoday.com/18847/life-of-the-sun/). For future studies, better-equipped space telescopes, with biosignature detectors, are planned, such as the ESA's planetary transits and oscillations of stars, or NASA's James Webber space telescope (http://sci.esa.int/jump.cfm?oid=42277), which will give us further scientific parameters with which to search for life outside our solar system.

## 5. References

Baumann H (1922) Die Anabiose der Tardigraden. *ZoologischeJahrbücher***45**, 501–556.

Borgonie G, García-Moyano A,Litthauer D, Bert W, Bester A, van Heerden E, Möller C, Erasmus M and OnstottTC (2011) Nematoda from the terrestrial deep subsurface of South Africa. *Nature***474**, 79–82. doi:10.1038/nature09974.

Creer BY, Smedal CHA and Wingrove RC (2012) Centrifuge Study of Pilot Tolerance to acceleration and the effects of acceleration on pilot performance.*NASA Technical note* D-337.

Clegg JS (1973) Do dried cryptobiotes have a metabolism? In:*Anhydrobiosis. Dowden, Hutchinson and Ross, Stroudsburg*, 141–147.

Degma P andGuidetti R (2007) Notes to the current checklist of Tardigrada. *Zootaxa***1579**, 41–53.

Degma P, Bertolani R and GuidettiR (2009-2017) Actual checklist of Tardigrada species. (Ver. 33: 15-10-2017).

http://www.tardigrada.modena.unimo.it/miscellanea/Actual%20checklist%20of%20Tardigrada.pdf

Erdmann W and Kaczmarek Ł (2017) Tardigrades in Space Research - Past and Future.*Origins of Life and Evolution of Biospheres***47**, 545–553.

Förster F, Liang C, Shkumatov A, Beisser D, Engelmann JC, Schnölzer M, Frohme M, Müller T, Schill RO and Dandekar T (2009) Tardigrade workbench: comparing stress-related proteins, sequence-similar and functional protein clusters as well as RNA

elements in tardigrades. *BMC Genomics***10**, 469. pmid:19821996.doi: 10.1186/1471-2164-10-469.

Grohme MA, Mali B, Wełnicz W, Michel S, Schill RO andFrohme M (2013) The aquaporin channel repertoire of the tardigrade *Milnesium tardigradum*. *BioinformBiol Insights***7**, 153–165. doi: 10.4137/BBI.S11497.

Guidetti R andBertolani R (2005) Tardigrade taxonomy: an updated check list of the taxa and a list of characters for their identification. *Zootaxa***845**, 1–46.

Guidetti R, Altiero T andRebecchi L (2011) On dormancy strategies in tardigrades. *Journal of Insect Physiology***57**, 567–576.

Guidetti R, Rizzo AM, Altiero T andRebecchi L (2012) What can we learn from the toughest animals of the Earth? Water bears (tardigrades) as multicellular model organisms in order to perform scientific preparations for lunar exploration. *Planetary and Space Science***74**, 97–102.

Hashimoto T, Horikawa DD, Saito Y, Kuwahara H, Kozuka-Hata H, Shin-I T, Minakuchi Y, Ohishi K, Motoyama A, Aizu T, Enomoto A, Kondo K, Tanaka S, Hara Y, Koshikawa S, Sagara H, Miura T, Yokobori S, Miyagawa K, Suzuki Y, Kubo T, Oyama M, Kohara Y, Fujiyama A, Arakawa K, Katayama T, Toyoda A and Kunieda T (2016) Extremotolerant tardigrade genome and improved radiotolerance of human cultured cells by tardigrade-unique protein. *Nature Communications***7**, 12808.

Hengherr S, Heyer AG, Köhler HR andSchill RO (2007)Trehalose and anhydrobiosis in tardigrades - evidence for divergence in response to dehydration. *The FEBS Journal***275**, 281–288.

Hengherr S, Worland MR, Reuner A, Brummer F andSchill RO (2009) Freeze tolerance, supercooling points and ice formation: comparative studies on the subzero temperature survival of limno-terrestrial tardigrades. *Journal of Experimental Biology***212**, 802–807.

Horikawa DD (2008) The tardigrade *Ramazzottiusvarieornatus* as a model animal for astrobiological studies. *Biological Sciences in Space***22**, 93–98.

Horikawa DD, Sakashita T, Katagiri C, Watanabe M, Kikawada T, Nakahara Y, Hamada N, Wada S, Funayama T, Higashi S, Kobayashi Y, Okuda T andKuwabara M (2006) Radiation tolerance in the tardigrade *Milnesium tardigradum*. *International Journal of Radiation Biology***82**, 843–848.

Jönsson KI (2007) Tardigrades as a potential model organism in space rsearch. *Astrobiology***7**, 757–766.

Jönsson KI andSchill RO (2007) Induction of Hsp70 by desiccation, ionising radiation and heat-shock in the eutardigrade *Richtersiuscoronifer*. *Comparative Biochemistry and Physiology Part B: Biochemistry and Molecular Biology* **146**, 456–460.

Jönsson KI andGuidetti R (2001) Effects of methyl-bromide fumigation on anhydrobiotic micrometazoans. *Ecotoxicology and Environmental Safety***50**, 72–75.

Jönsson KI, Rabbow E, Schill RO, Harms-Ringdahl M andRettberg P (2008) Tardigrades survive exposure to space in low Earth orbit. *Current Biology***18**, R729–R731.

Kashyap JM and Kaczmarek Ł (2017) ATI and CTI, Mendeley Data, v1.http://dx.doi.org/10.17632/d3fp4wsmx3.1.

Kashyap JM, GudennavarSB, Doshi U andSafonova M (2017) Indexing of exoplanets in search for potential habitability: application to Mars-like worlds.*Astrophysics and Space Science***362**, 146.

LeukoS, Rettberg P, Pontifex AL and Burns BP (2014) On the response of halophilic Archaea to space conditions.*Life***4**, 66–76. doi:10.3390/life4010066.

Mayor M. and Queloz D (1995) A Jupiter-mass companion to a solar-type star. *Nature* **378**, 355.

McInnes SJ (1994) Zoogeographic distribution of terrestrial /freshwater tardigrades from current literature. *Journal of Natural History* **28**, 257–352.

Nelson DR, Guidetti R andRebecchi L (2010) Chapter 14: Tardigrada. In *Ecology and classification of North American Freshwater Invertebrates. 3rd ed.* San Diego: Academic Press, 455–484.

Nelson DR, Guidetti R andRebecchi L (2015) Chapter 17: Phylum Tardigrada. In *Ecology and General Biology: Vol. 1: Thorp and Covich's Freshwater Invertebrates* 347–380.

Ono F, Saigusa M, Uozumi T, Matsushima Y, Ikeda H, Saini NL and Yamashita M (2008) Effect of high hydrostatic pressure on to life of the tiny animal tardigrade. *Journal of Physics and Chemistry of Solids* **69**, 2297–2300.

Oren A (2014) Halophilic archaea on Earth and in space: growth and survival under extreme conditions. *Philosophical Transactions of the Royal Society A* **372(2030)**, pii: 20140194.doi: 10.1098/rsta.2014.0194.

Pigoń A andWęglarska B (1955) Rate of metabolism in tardigrades during active life and anabiosis. *Nature* **176**, 121–122.

Ramazzotti G andMaucci W (1983) Il Phylum Tardigrada. *Memorie dellí IstitutoItaliano di Idrobiologia, Pallanza* **41**, 1–1012.

Ramløv H andWesth P (1992) Survival of the cryptobiotic eutardigrade *Adorybiotuscoronifer* during cooling to minus 196˚C: effect of cooling rate, trehalose level, and short-term acclimation. *Cryobiology* **29**, 125–130.

Ramløv H andWesth P (2001) Cryptobiosis in the eutardigrade *Adorybiotus* (*Richtersius*) *coronifer*: tolerance to alcohols, temperature and de novo protein synthesis. *ZoologischerAnzeiger* **240**, 517–523.

Rampelotto PH (2013) Extremophiles and Extreme Environments. *Life* **3**, 482–485. doi:10.3390/life3030482.

Rizzo AM, Negroni M, Altiero T, Montorfano G, Corsetto P, Berselli P, Berra B, Guidetti R and Rebecchi L (2010) Antioxidant defences in hydratedand desiccated states of the tardigrade *Paramacrobiotusrichtersi*. *Comparative Biochemistry and Physiology Part B: Biochemistry and Molecular Biology* **156**, 115–121.

Sancho LG, de la Torre R, Horneck G, Ascaso C, de Los Rios A, Pintado A, Wierzchos J and Schuster M (2007) Lichens survive in space: results from the 2005 LICHENS experiment. *Astrobiology* **7**, 443–454.

Schulze-Makuch D, Méndez A, Fairén AG, von Paris P, Turse C, Boyer G, Davila AF, António MR, Catling Dand Irwin LN (2011) A two-tiered approach to assessing the habitability of exoplanets". *Astrobiology* **11**, 1041–1052.

Seckbach J (2000)Extremophilies as Models for Extraterrestrial Life. *Bioastronomy 99: A New Era in the Search for Life, ASP Conference Series* **213**, 379.

Seki K and Toyoshima M (1998) Preserving tardigrades under pressure. *Nature* **395**, 853–858.

Sengupta S (2016) The search for another earth. *Resonance* **21**, 641–652.

Wełnicz W, Grohme MA, Kaczmarek Ł, Schill RO andFrohme M (2011) Anhydrobiosis in tardigrades - the last decade. *Journal of Insect Physiology* **57**, 577–583.

Wright JC (2001) Cryptobiosis 300 years on from van Leeuwenhoek: what have we learned about tardigrades? *ZoologischerAnzeiger* **240**, 563–582.

Yamaguchi A, Tanaka S, Yamaguchi S, Kuwahara H, Takamura C, Imajoh-Ohmi S, Horikawa DD, Toyoda A, Katayama T, Arakawa K, Fujiyama A, Kubo T andKunieda T (2012)Two novel Heat-Soluble protein families abundantly expressed in an anhydrobiotic tardigrade. *PLoS ONE* **7**, e44209. doi:10.1371/journal.pone.0044209.